\documentclass[prb,aps,twocolumn,floatfix]{revtex4}
\usepackage{bm}
\usepackage{amssymb}
\usepackage{graphicx}

\begin{document}

\title{
Wavefunctions for large systems as basis for electronic structure calculations
}

\author{P. Fulde}
\affiliation{Max-Planck-Institut f\"ur Physik komplexer Systeme, 01187 Dresden, Germany}

\date{\today } 

\begin{abstract}
Electronic structure calculations for solids based on many-electron wavefunctions have been hampered by the argument that for large electron numbers wavefunctions are not a legitimate scientific concept, because they face an {\it exponential wall} problem. We show that this problem can be avoided by using cumulant techniques in formulating wavefunctions. Therefore calculations for solids based on many-electron functions are possible and useful. This includes also systems with strongly correlated electrons.
\end{abstract}
\maketitle

Electronic-structure calculations are one of the most important and active fields not only in condensed matter physics but also in chemistry from where they originate. They began shortly after Heisenberg \cite{Heisenberg25} and Schr\"odinger \cite{Schrodinger26} developed the quantum mechanical tools for treating atomic systems. The first molecule which was treated was naturally $H_2$ with two electrons only. Two approximations were applied which differ in the way the mutual electron repulsion is treated. In the historically first treatment of Heitler and London \cite{Heitler27} electron repulsions are considered more important than their kinetic energy and therefore the quantum mechanical wavefunction does not contain any ionic configurations $H^+ - H^-$ but only $H-H$ ones. This is what we would call now the strong correlation limit. This line of approximations was later extended by Pauling's resonating bond theory \cite{Pauling60} or Anderson's approach to the high-temperature superconducting cuprates \cite{Anderson87}. Also the Zhang-Rice singlet \cite{Zhang88} in these cuprates belongs to that category.

The second approach due to Hund \cite{Hund26} and Mulliken \cite{Mulliken28} treats the electron repulsions as being secondary and correspondingly in a mean-field approximation. The two electrons occupy a molecular orbital and the weight of ionic configuration $H^+ - H^-$ is as large as that of neutral atoms $H-H$. This approach was generalized by the Hartree-Fock theory \cite{Hartree28,Fock30} and has found rich applications in chemistry and solids both, on an ab initio and  on a semiempirical level. In Hartree-Fock approximation the electrons remain uncorrelated. In reality a system is always between the Heitler-London and Hund-Mulliken limit and therefore the question arises how to interpolate between the two limits. One way is to improve a self-consistent field (SCF) wavefunction by adding configurational interaction (CI) terms to it. This has been a highly accurate method employed in quantum chemistry for the treatment of small molecules \cite{Boys50}. However, for solids density-functional theory (DFT) \cite{Hohenberg64} has triggered a revolution in electronic structure theory, in particular since it could be combined with linearized methods for solving the Kohn-Sham equations \cite{Kohn65,Andersen75}, the heart of DFT. This success has also made by now DFT the most applied method in chemistry too.

The Kohn-Sham equations and their use show that it is basically a mean-field theory. However, it contains effects of electron correlations by the choice of the exchange-correlation potential $v_{xc} ({\bf r})$. It does not make any statement about the form of the many-electron wavefunction. Thus it is very different from a MO or HF theory. Despite this enormous success, DFT has also some weaknesses, which are difficult to correct. A general one is that its results depends on the choice of the exchange-correlation potential $v_{xc} ({\bf r})$. Thus any approximation to it is essentially uncontrolled and difficult to improve when it does not work very well. This is, e.g., the case for strongly correlated electron systems like the ones with heavy quasiparticles \cite{Zwicknagl92}. Although it is a theory for the ground state, the Kohn-Sham equations have also been applied to bandstructures. Here well known problems with the size of band gaps do appear \cite{Fulde95}. 

This suggests to develop in parallel with DFT also electronic structure calculations based on many-body wavefunctions. However, this development has been hampered by an argument put forward that the many-electron wavefunction $\psi \left( {\bf r}_1 \sigma_1, \dots, {\bf r}_N \sigma_N \right)$ for a system of $N$ electrons is not a legitimate scientific concept, when $N \geq N_0$ where $N_0 \approx 10^3$. \cite{Kohn99} With legitimacy it is meant in this context that (a) one must be able to calculate $\psi \left( {\bf r}_1 \sigma_1, \dots, {\bf r}_N \sigma_N \right)$ with sufficient accuracy and (b) that the wavefunction can be recorded sufficiently well so that it can be reproduced at later times.

The argument brought forward for (a) is that the overlap of any calculated ground-state wavefunction $\psi_{\rm cal}$ with the exact one $\psi_0$ is $|\langle \psi_0 | \psi_{\rm cal} \rangle | = (1 - \epsilon)^N$ and therefore goes to zero when $N \to \infty$. Here $\epsilon$ is the minimum error one has to deal with, when for a small system the exact wavefunction is approximated by a calculated one. The nearly vanishing overlap is considered as evidence, that the many-body wavefunction cannot be calculated with sufficient accuracy. This argument will be scrutinized. Concerning (b) it is noticed that when it takes $m$ bits to describe one electron, the total number of bits grows as $m^N$ for an $N$ electron system. Thus it cannot be recorded for large values of $N$. The exponential decrease with $N$ of the overlap of an approximate with the exact wavefunction and the exponential growth of bits required to record the many-electron wavefunction are called the {\it exponential wall} one is facing when using many-electron wavefunctions.

Before we discuss the elimination of the exponential wall we recall the exponential redundancy of information which the wavefunction of the above type has. Assume that we deal with $N$ non-interacting atoms, e.g., He atoms, with a single two-particle excitation added to the SCF ground state for its improvement. In this case the total wavefunction contains $2^N$ terms. However, the only relevant information is the actual amplitude of the admixed two-particle excitation to the SCF ground state, i.e., the wavefunction of a single He atom. Therefore we must search for a representation in which only the important information is contained with the remaining redundancy eliminated. This is achieved by employing cumulants.

Cumulants have played a role in physics and chemistry for a long time. Kubo \cite{Kubo62} explored and demonstrated their usefulness in various branches of physics and chemistry. With their help he gave, e.g., an elegant derivation of the Ursell-Mayer cluster expansion \cite{Mayer40} for classical and quantum gases. He pointed out that``generalized cumulant expansions provide us with a point of view from which many existent methods in quantum mechanics and statistical mechanics can be unified''. We use them here in order to express the many-electron wavefunction in a form which is free of the exponential wall and therefore well suited for electronic structure calculations for large systems including solids. This is possible since cumulants deal only with statistically connected processes and eliminate this way exponential redundancies of information.

\section*{Cumulant wave operator}

For the definition of cumulants and some of their properties we refer to the literature \cite{Kubo62,Fulde95,Kladko98}. Here it suffices to write down the cumulant of the product of two operators $A_1 A_2$ formed with two vectors in Hilbert space $\phi_1$ and $\phi_2$ with nonzero overlap $\langle \phi_1 | \phi_2 \rangle \neq 0$,

\begin{equation}
\label{Eq1}
\langle \phi_1 | A_1 A_2 | \phi_2 \rangle^c = \frac{\langle \phi_1 | A_1 A_2 | \phi_2 \rangle}{\langle \phi_1 | \phi_2 \rangle} - \frac{\langle \phi_1 | A_1 | \phi_2 \rangle}{\langle \phi_1 | \phi_2 \rangle} \times \frac{\langle \phi_1 | A_2 | \phi_2 \rangle}{\langle \phi_1 | \phi_2 \rangle}~~.
\end{equation}

\noindent Furthermore $\langle \phi_1 | A | \phi_2 \rangle^c = \langle \phi_1 | A | \phi_2 \rangle / \langle \phi_1 | \phi_2\rangle$ and $\langle \phi_1 | 1 | \phi_2 \rangle^c = \ln \langle \phi_1 | \phi_2\rangle$. Eq. (\ref{Eq1}) demonstrates the most important feature which is that the cumulant of a matrix element keeps only those contributions which cannot be factorized. We want to transform the vector $\phi_2$ into another vector $\psi$ by a sequence of infinitesimal transformations $\phi'_2 = e^{\delta S} \phi_2 = (1 + \delta S) \phi_2$ with $\delta S = \epsilon S$, $\epsilon \ll 1$. Then an expression of the form $\langle \phi_1 | A | \phi_2 \rangle$ transforms into

\begin{equation}
\label{Eq2}
\langle \phi_1 | A | \psi \rangle = \langle \phi_1 | A \Omega | \phi_2 \rangle^c
\end{equation}

\noindent with $\Omega = \lim\limits_{L \to \infty} \prod\limits^L_{i=1} (1 + \delta S_i)$. A derivation of this expression is found in Ref. \onlinecite{Kladko98}. We apply these findings to electronic wavefunctions by identifying $| \phi_1 \rangle$ with the SCF ground state $| \Phi_0 \rangle$ and $| \psi \rangle$ with the wavefunction $| \psi_0 \rangle$ of the exact ground state. Conventionally a wave- or Moeller operator $\tilde{\Omega}$ is defined which connects the two, i.e., $| \psi_0 \rangle = \tilde{\Omega} | \Phi_0 \rangle$. Therefore we call $\Omega$ in Eq. (\ref{Eq2}) which transforms $\Phi_0$ into $\psi_0$ the cumulant wave operator. It is a vector in operator space with the metric

\begin{equation}
\label{Eq3}
(A | B) = \langle\Phi_0 | A^+ B | \Phi_0 \rangle^c~~.
\end{equation}

\noindent When $A$ is identified with the Hamiltonian $H$ we obtain immediately for the ground-state energy

\begin{eqnarray}
\label{Eq4}
E_0 & = & \langle \Phi_0 | H \Omega | \Phi_0 \rangle^c \nonumber \\[2ex]
& = & (H | \Omega)~~.
\end{eqnarray}

\noindent Note that $| \Omega)$ is not unique since the transformation from $| \Phi_0 \rangle$ to $| \psi_0 \rangle$ may take place along different paths in Hilbert space. Yet, independent of any choice of $| \Omega)$ all cumulants calculated with it are independent of the chosen path.

\section*{Avoiding the exponential wall}

Instead of characterizing the ground-state wavefunction of an electron system by a vector $| \psi_0 \rangle$ in Hilbert space we define it by the cumulant wave operator $| \Omega )$ in operator space with the (cumulant) metric (\ref{Eq3}). We sum the infinitesimal transformations discussed before and write

\begin{equation}
\label{Eq5}
| \Omega) = | 1 + S )
\end{equation}

\noindent where $| S )$ is the cumulant scattering matrix. As shown in Ref. \onlinecite{Kladko98} it has the form of a resolvent. When $H$ is divided into a SCF and a residual interaction part, i.e., $H = H_{\rm SCF} + H_{\rm res}$ this form is

\begin{equation}
\label{Eq6}
| S ) = \lim\limits_{z \to 0} \left| \frac{1}{z - H} H_{\rm res} \right)  ~~.
\end{equation}

\noindent Thus $S$ can be expanded in powers of $H_{\rm res}$ if required. A powerful tool for the determination of $| S )$ is the use of the decomposition

\begin{equation}
\label{Eq7}
| S ) = \left| \sum_I S_I +  \sum_{\langle IJ \rangle} \delta S_{IJ} + \sum_{\langle IJK \rangle} \delta S_{IJK} + \dots \right) 
\end{equation}

\noindent with site indices $IKL$. Furthermore $\delta S_{IJ} = S_{IJ} - S_I - S_J$ and correspondingly for $\delta S_{IJK}$. The different contributions to $| S )$ involve only a small number of electrons each. 

Before we discuss their determination we want first to point out that by the use of cumulants the {\it exponential walls} do not appear and therefore don't pose any problem. The following relations hold between the vectors $| \psi_0 \rangle$ and $| \Phi_0 \rangle$ in Hilbert space and the cumulant waveoperator $\Omega$

\begin{eqnarray}
\label{Eq8}
| \psi_{0, {\rm norm}} \rangle & = & \frac{| \psi_0 \rangle}{\langle\Phi_0 | \psi_o \rangle} \nonumber \\[2ex]
& = & \nabla _{\Phi_0} \langle \Phi_0 | 1 | \psi_0 \rangle^c \nonumber \\[2ex]
& = & \nabla _{{\rm left} \Phi_0} \langle \Phi_0 | \Omega | \Phi_0 \rangle^c~~.
\end{eqnarray}

\noindent Since for a cumulant $\langle \phi_1 | A_1 A_2 | \alpha \phi_2 \rangle^c = \langle \phi_1 | A_1 A_2 | \phi_2 \rangle^c$, differences in $\langle \Phi_0 | \psi_0 \rangle$ when approximations for $| \psi_0 \rangle$ are being made, are unimportant. They don't appear in cumulant representation. This removes the exponential wall problem (a).

Next we have to show that $| S )$ can be calculated with sufficient accuracy and can be well recorded, i.e., it does not require an exponential increase in the number of bits as $N$ increases.

\section*{Evaluation and application of the cumulant scattering matrix}

First we consider the evaluation of the one-site scattering matrix $S_I$. We express $| \Phi_0 \rangle$ in terms of occupied localized (Wannier) orbitals. The latter are usually expressed in Gaussian type orbitals (GTO's) forming the basis set. Their number depends on the desired degree of accuracy of the results. All electrons in these Wannier orbitals are kept frozen except those in orbitals centered at site $I$. These are correlated among each other and the number of configurations included depends on the size of the basis set.

For strongly correlated electrons we can identify $S_I$ with complete active space SCF (CASSCF) calculations for electrons on that site. This provides for a suitable way of treating strongly correlated electrons in a solid with quantum chemical methods. We do not know of an alternative way of setting up CAS-SCF calculations for an infinite solid. Each term $S_I$ contributes according to (\ref{Eq4}) an increment $\epsilon_I$ to the correlation energy. Dealing hereby with cumulants poses no problem. When matrix elements are computed one merely has to consider completely linked contractions of operators only and to discard all other contributions. Similarly $S_{IJ}$ is calculated by keeping all electrons in Wannier orbitals frozen except those centered at sites $I$ and $J$. They are simultaneously correlated and yield a contribution $\epsilon_{IJ}$ to the correlation energy. The method of increments \cite{Stoll92} is a rather powerful and accurate scheme \cite{Stoll05}. It is usually fast convergent as long as the system has an excitation gap. Already terms $\delta S_{IJK}$ give usually small corrections to the correlation energy. Metals require special attention because Wannier orbitals for partially filled bands are not well localized \cite{Kohn73}. More details concerning $| S )$ are found in \cite{Fulde12}. 

The point we want to emphasize is that a documentation of the results for the different contributions to $| S )$ poses no problem. Only a relatively small number of electrons is involved in each increment and the number of basis functions, i.e., GTO's for a given site and its surroundings is small. It does not change when the total electron number $N$ is increased. There is no exponential bit increase with increasing electron number because all redundancies have been removed. Thus there is no principle obstacle which prevents us from many-body wavefunction based electronic structure calculations. In fact, a number of systems have already been studied by using the method of increments and results, e.g., for the binding energy, force constant et cetera \cite{Fulde12,Paulus10,Schuetz99}. They have been compared favourably with DFT results \cite{Paulus10}. The incremental decomposition of the energy resembles the Bethe-Goldstone expansion \cite{Bethe57} from nuclear physics. Yet, the determination of the many-body wavefunction stressed here is hardly possible in this approach. Because of the exponential wall argument these calculations have been considered as justified only for relatively small clusters. As shown here this is not the case.

Summarizing we reemphasize that by defining the many-electron wavefunction through the cumulant wave operator $| \Omega )$ instead of a vector $| \psi_0 \rangle$ in Hilbert space we avoid the exponential wall connected with the latter. This removes a hindrance for wavefunction based electronic structure calculations of solids. It also allows for including solids with strongly correlated electrons in this scheme.

\section*{Acknowledgement}

I would like to thank K. Kladko for extensive discussions on cumulants and H. Stoll for a long standing cooperation on the subject of this paper.

\end{document}